\documentclass[10pt,letterpaper]{article}

\usepackage[verbose=true,letterpaper]{geometry}

\usepackage[utf8]{inputenc} 
\usepackage[T1]{fontenc}    
\usepackage{amsmath}
\usepackage{epsfig}
\usepackage{graphicx}
\usepackage{wrapfig}
\usepackage{psfrag}
\usepackage{substr}
\usepackage[tight]{minitoc}
\usepackage{subfigure}
\usepackage{longtable}
\usepackage{chngcntr}
\usepackage[all]{xy}
\usepackage{cite}
\usepackage{xr}

\providecommand{\keywords}[1]
{
  \small	
  \textbf{\textit{Keywords---}} #1
}

\title{Dynamics of phases and chaos in models of locally coupled conservative or dissipative oscillators}

\author{
  Vyacheslav P.~Kruglov \\
  Saratov Branch Kotelnikov Institute of Radioengineering and Electronics of RAS\\
  Saratov, Russia\\
  Yuri Gagarin State Technical University\\
  Saratov, Russia\\
  \texttt{kruglovyacheslav@gmail.com} \\
   \and
 Sergey P.~Kuznetsov\\
  Saratov Branch Kotelnikov Institute of Radioengineering and Electronics of RAS\\
  Saratov, Russia\\
  Udmurt State University\\
  Izhevsk, Russia\\
  \texttt{spkuz@yandex.ru} \\
}

\begin{document}
\maketitle

\begin{abstract}
We discuss Hamiltonian model of oscillator lattice with local coupling. Model describes spatial modes of nonlinear Schr\"{o}dinger
equation with periodic tilted potential. The Hamiltonian system manifests reversibility of Topaj -- Pikovsky phase oscillator lattice. 
Furthermore, the Hamiltonian system has invariant manifolds with dynamics exactly equivalent to the Topaj -- Pikovsky model. We demonstrate the complexity
of dynamics with results of numerical simulations. We also propose two dissipative models close to Topaj -- Pikovsky system.
\end{abstract}


\section{Introduction}

Topaj and Pikovsky introduced~\cite{1} a lattice of $N$ locally coupled phase oscillators:
\begin{equation}
\dot{\phi} _j = \omega _j + \varepsilon \sin \left( \phi _{j+1} - \phi _j \right)  + \varepsilon \sin \left( \phi _{j-1} - \phi _j \right),
\label{eq1}
\end{equation}
where $\phi _j$ are phases of oscillators, $\omega _j$ are frequencies of oscillators $j = 1, \dots, N$, $\varepsilon$ are coupling constants, the $\sin \Delta\phi$  provides coupling between two neighboring oscillators and depends
only on phase differences. Instead of $\sin \Delta\phi$ one can choose different $2\pi$-periodic odd function. Boundary conditions are $\phi _0 = \phi _1$ and $\phi _{N+1} = \phi _N$, so the first and the last oscillators are not fixed.

We simplify system of equations~\eqref{eq1} since their right-hand sides depend only on phase differences:
\begin{equation}
\dot{\psi} _j = \Delta _j + \varepsilon \sin \psi _{j+1}  + \varepsilon \sin \psi _{j-1}
- 2 \varepsilon \sin \psi _{j},
\label{eq2}
\end{equation}
where $\psi _j = \phi _{j+1} - \phi _{j} $ are phase shifts between two neighboring oscillators, $\Delta _j = \omega _{j+1} - \omega _j $ are frequency shifts. The number of variables is $N-1$. Boundary conditions are 
$\psi _0 = 0$ and $\psi _N = 0$. 

Topaj and Pikovsky showed that if $\varepsilon \ll 1 $ and frequencies are far from resonant, then system~\eqref{eq2} demonstrates quasiperiodic motions~\cite{1}. At large $\varepsilon$ a fully phase-locked state is 
observed~\cite{1}. At intermediate values of $\varepsilon$ phase-locked clusters of oscillators emerge. A scenario of transition from the completely phase-locked state to the quasi-periodic state depends on the frequencies
$\omega _j$. Topaj and Pikovsky chose frequencies linearly distributed such that $\Delta _j = 1$. With that at small $\varepsilon$ the average divergence of vector field given by right-hand sides of equations~\eqref{eq2} is
close to zero~\cite{1}. This means that dynamics of lattice at small $\varepsilon$ is close to conservative. 
 
The observed in~\eqref{eq2} quasi-conservative behavior appears due to reversibility of the dynamics~\cite{1,2}. A dynamical system is reversible if there is an involution in phase space which reverses the direction of
time~\cite{3}.
Involution is a transformation $\mathbf{R}$ that, if composed with itself yields the identity. Reversible equations are invariant under the combined action of the involution and time reversal. An involution that achieves this is often
called a reversing symmetry of the system. Trajectories that are invariant by involution are called symmetric~\cite{3}. These symmetric trajectories have inverse Lyapunov spectra, because the Lyapunov exponents change sign with
the time inversion~\cite{1}. Symmetric trajectories may be periodic, non-wandering (quasi-periodic or chaotic) or heteroclinic, connecting symmetric attractor and repeller. The latter is possible in dissipative systems~\cite{3}
and systems with nonholonomic constraints~\cite{4}. 

For equations~\eqref{eq2} involution is 
\begin{equation}
\mathbf{R}: \psi _j \mapsto \pi - \psi _{N-j}.
\label{eq3}
\end{equation}
Involution~\eqref{eq3} has an invariant set $\text{Fix} \mathbf{R}: \psi _j + \psi _{N-j} = \pi$. Trajectories crossing it are symmetric. Topaj and Pikovsky showed numerically in~\cite{1} the reversibility of~\eqref{eq2} for
trajectories starting from invariant set $\text{Fix} \mathbf{R}$. 

The aim of this work is to observe the phenomenon of reversibility in more general lattices of oscillators. We start with Hamiltonian lattice of oscillators describing Bose -- Einstein condensate in a tilted periodic potential
~\cite{5}. Dynamics of this lattice is very similar to the Topaj -- Pikovsky model.

\vspace{-1mm}

\section{Hamiltonian lattice model}

It was pointed out in~\cite{6} that phase space of the Hamiltonian model proposed in~\cite{5} includes invariant manifolds with dynamics the same as~\eqref{eq1}. We start with Hamiltonian function
\begin{equation}
\begin{aligned}
\mathcal{H} \left( \ldots , z_j, \bar{z}_j, \ldots \right) & = 
\sum_{j=1}^{N} \omega _j z_j \bar{z}_j + \frac{1}{2} \beta \sum_{j=1}^{N} z_j^2 \bar{z}_j^2 + 
\\ & + \frac{i \varepsilon}{2} \sum_{j=1}^{N} \left( z_{j+1} \bar{z}_{j+1} - z_j \bar{z}_j \right) \left( z_{j+1} \bar{z}_j - z_j \bar{z}_{j+1} \right) +
\\ & + \frac{i \varepsilon}{2} \sum_{j=1}^{N} \left( z_{j-1} \bar{z}_{j-1} - z_j \bar{z}_j \right) \left( z_{j-1} \bar{z}_j - z_j \bar{z}_{j-1} \right),
\label{eq4}
\end{aligned}
\end{equation}
where $z_j$ are complex amplitudes of Wannier -- Stark resonant states $u\left(x,t\right)=\sum_{j=1}^{N} z_j\left(t\right) w_j\left(x\right)$ of nonlinear Schr\"{o}dinger equation with Hamiltonian~\cite{5}:
\begin{equation}
\begin{aligned}
\Hat{\mathcal{H}} = - \frac{\hbar^2}{2m}\partial_{xx} + V_0\cos^2\left(k x\right) +F x + g|u|^2.
\label{eq5}
\end{aligned}
\end{equation}
Wannier -- Stark states are assumed localized which leads to local coupling in~\eqref{eq4}. Frequencies in~\eqref{eq4} are equidistant: $\omega _j = - \frac{\pi F}{k \hbar} j$. 

One can decompose $z_j$ to real and imaginary parts, which are in fact momentums and coordinates of oscillators: $z_j = \frac{1}{\sqrt{2}} \left(p_j + i q_j\right) $, or to actions (intensities of oscillations or populations of
potential wells) and angles (phases of oscillations): 
$z_j = \sqrt{I_j} e^{i \phi _j} $. Hamiltonian function~\eqref{eq4} can be rewritten in both kinds of real variables:
\begin{equation}
\begin{aligned}
\mathcal{H} \left( \ldots , p_j, q_j, \ldots \right) & = 
\frac{1}{2} \sum_{j=1}^{N} \omega _j \left( p_j^2 + q_j^2 \right) + \frac{1}{8} \beta \sum_{j=1}^{N} \left( p_j^2 + q_j^2 \right)^2 - 
\\ & - \frac{\varepsilon}{4} \sum_{j=1}^{N} \left( p_{j+1}^2 + q_{j+1}^2 - p_j^2 - q_j^2 \right) \left( q_{j+1} p_j - q_j p_{j+1} \right) -
\\ & - \frac{\varepsilon}{4} \sum_{j=1}^{N} \left( p_{j-1}^2 + q_{j-1}^2 - p_j^2 - q_j^2 \right) \left( q_{j-1} p_j - q_j p_{j-1} \right),
\label{eq6}
\end{aligned}
\end{equation}
\begin{equation}
\begin{aligned}
\mathcal{H} \left( \ldots , I_j, \phi_j, \ldots \right) & = 
\sum_{j=1}^{N} \omega _j I_j + \frac{1}{2} \beta \sum_{j=1}^{N} I_j^2 - 
\\ & - \varepsilon \sum_{j=1}^{N} \sqrt{I_{j+1} I_j} \left( I_{j+1} - I_j \right) \sin \left( \phi_{j+1} - \phi_j \right) -
\\ & - \varepsilon \sum_{j=1}^{N} \sqrt{I_{j-1} I_j} \left( I_{j-1} - I_j \right) \sin \left( \phi_{j-1} - \phi_j \right).
\label{eq7}
\end{aligned}
\end{equation}

Hamiltonian function~\eqref{eq7} produces equations
\begin{equation}
\begin{aligned}
\dot{I}_j = - \frac{\partial\mathcal{H}}{\partial\phi_j} = & - 2 \varepsilon \sqrt{I_{j+1} I_j} \left( I_{j+1} - I_j \right) \cos \left( \phi_{j+1} - \phi_j \right) - 
\\ & - 2 \varepsilon \sqrt{I_{j-1} I_j} \left( I_{j-1} - I_j \right) \cos \left( \phi_{j-1} - \phi_j \right),
\\ \dot{\phi}_j = \frac{\partial\mathcal{H}}{\partial I_j} = & \omega _j + \beta I_j + \varepsilon \left\{ 3 \sqrt{I_{j+1} I_j} - I_{j+1} \sqrt{\frac{I_{j+1}}{I_j}} \right\} \sin \left( \phi_{j+1} - \phi_j \right) +
\\ & + \varepsilon \left\{ 3 \sqrt{I_{j-1} I_j} - I_{j-1} \sqrt{\frac{I_{j-1}}{I_j}} \right\} \sin \left( \phi_{j-1} - \phi_j \right).  
\label{eq9}
\end{aligned}
\end{equation}
Boundary conditions are $\phi _0 = \phi _1$,  $\phi _{N+1} = \phi _N$, $I _0 = I _1$,  $I _{N+1} = I _N$.
If populations of all oscillators are equal to each other, $I_j = I$, they are constant, $\dot{I}_j = 0$. Therefore a family of invariant tori exists with constant equal populations $I$ of oscillators and phases $\phi_j$
governed by
\begin{equation}
\dot{\phi}_j = \omega _j + \beta I + 2 \varepsilon I \sin \left( \phi_{j+1} - \phi_j \right)
+ 2 \varepsilon I  \sin \left( \phi_{j-1} - \phi_j \right). 
\label{eq10}
\end{equation}
This is just the system~\eqref{eq1} with rescaled coupling $2 \varepsilon I$ and shifted frequencies $\omega _j + \beta I$. Thus, on the invariant torus $I_j = I$ the Hamiltonian function~\eqref{eq7} generates a system of coupled phase oscillators with local coupling. As before, one can change phases $\phi _j$ to phase shifts $\psi _j = \phi _{j+1} - \phi _j$ and reduce the phase space dimension in Hamiltonian model~\eqref{eq9} by one.
System~\eqref{eq10} is reversible, and Hamiltonian model~\eqref{eq9} is also reversible with involution
\begin{equation}
\begin{aligned}
\mathbf{R}: I_j \mapsto I_{N-j+1}, \
\psi _j \mapsto \pi - \psi _{N-j}.
\label{eq11}
\end{aligned}
\end{equation}

Involution~\eqref{eq11} has an invariant set $\text{Fix} \mathbf{R}: I_j = I_{N-j+1},\ \phi _{j+1} - \phi _j = \pi - (\phi _{N-j+1} - \phi _{N-j})$. 

System~\eqref{eq9} has two constants of motion~\cite{6}. One of them is the Hamiltonian function $\mathcal{H}$, another is the total population of oscillators (sum of intensities):
\begin{equation}
C^2 = \sum_{j=1}^{N} I_j = \sum_{j=1}^{N} z_j \bar{z}_j = \frac{1}{2} \sum_{j=1}^{N} \left( p_j^2 + q_j^2 \right).
\label{eq12}
\end{equation}
$C^2$ is constant due to norm preservation of nonlinear Schr\"{o}dinger equation~\eqref{eq5}. It makes the dynamics equivariant under a simultaneous scaling $\left( p_j,\ q_j \right) \mapsto \left( C p_j,\  Cq_j \right)$ and 
parameter transformation $\varepsilon \mapsto \varepsilon / C^2 $, $\beta \mapsto \beta / C^2 $ for all $j$ and any $C>0$ \cite{6}. Therefore we fix the normalization $C^2 = N/2$, where $N$ is the number of oscillators, without
loss of generality, and thus we define invariant tori with phase dynamics by setting populations locked $I_j = 1/2$. The dynamics is also invariant with respect to a global phase shift, because equations~\eqref{eq9}
depend only on the phase differences. Two constants of motion and the phase shift invariance make the phase space of the full Hamiltonian system effectively $(2N-3)$-dimensional. 

\section{Results of numerical simulation}

Equations~\eqref{eq9} for lattices of $N=3$ and $4$ oscillators were solved numerically with Runge -- Kutta $4^{th}$-order method. Simulations were ran with check of constants of motions preservation up to numerical errors. We compared numerical solutions
of the equations~\eqref{eq9} with results for phase lattice from Topaj and Pikovsky work~\cite{1}. 

\begin{figure}[!h]
\includegraphics[width=.5\textwidth,keepaspectratio]{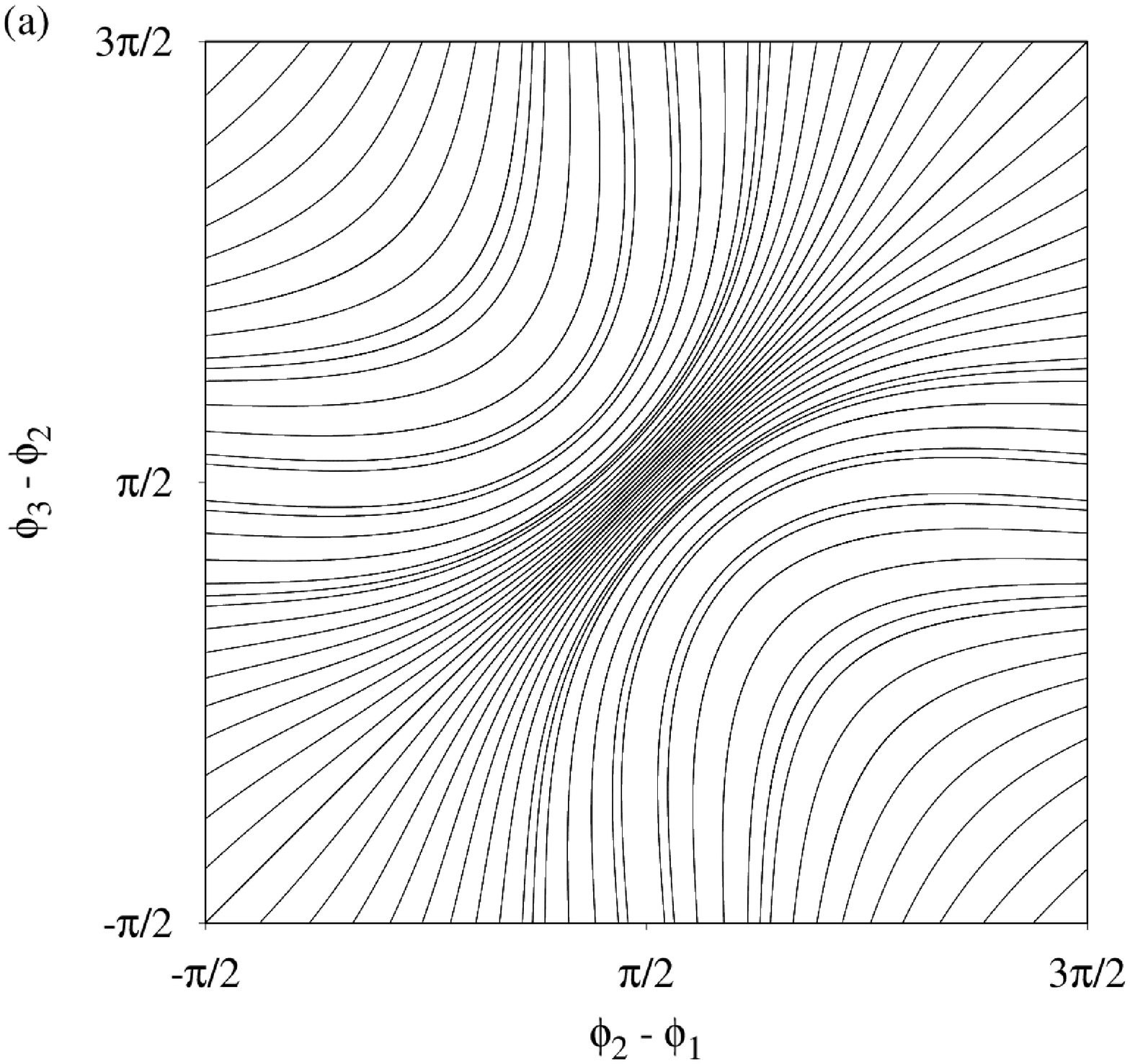}
\includegraphics[width=.5\textwidth,keepaspectratio]{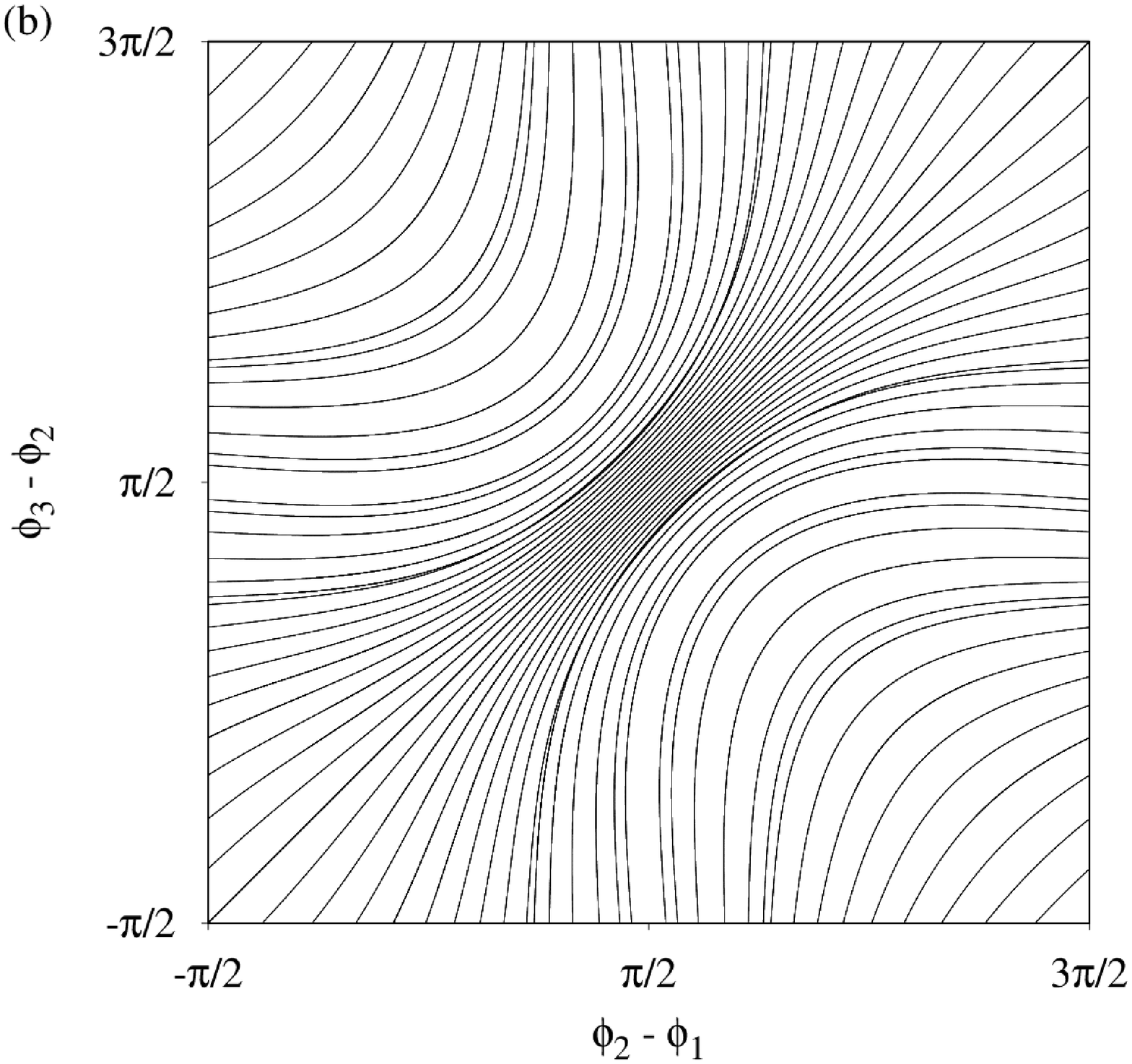}
\includegraphics[width=.5\textwidth,keepaspectratio]{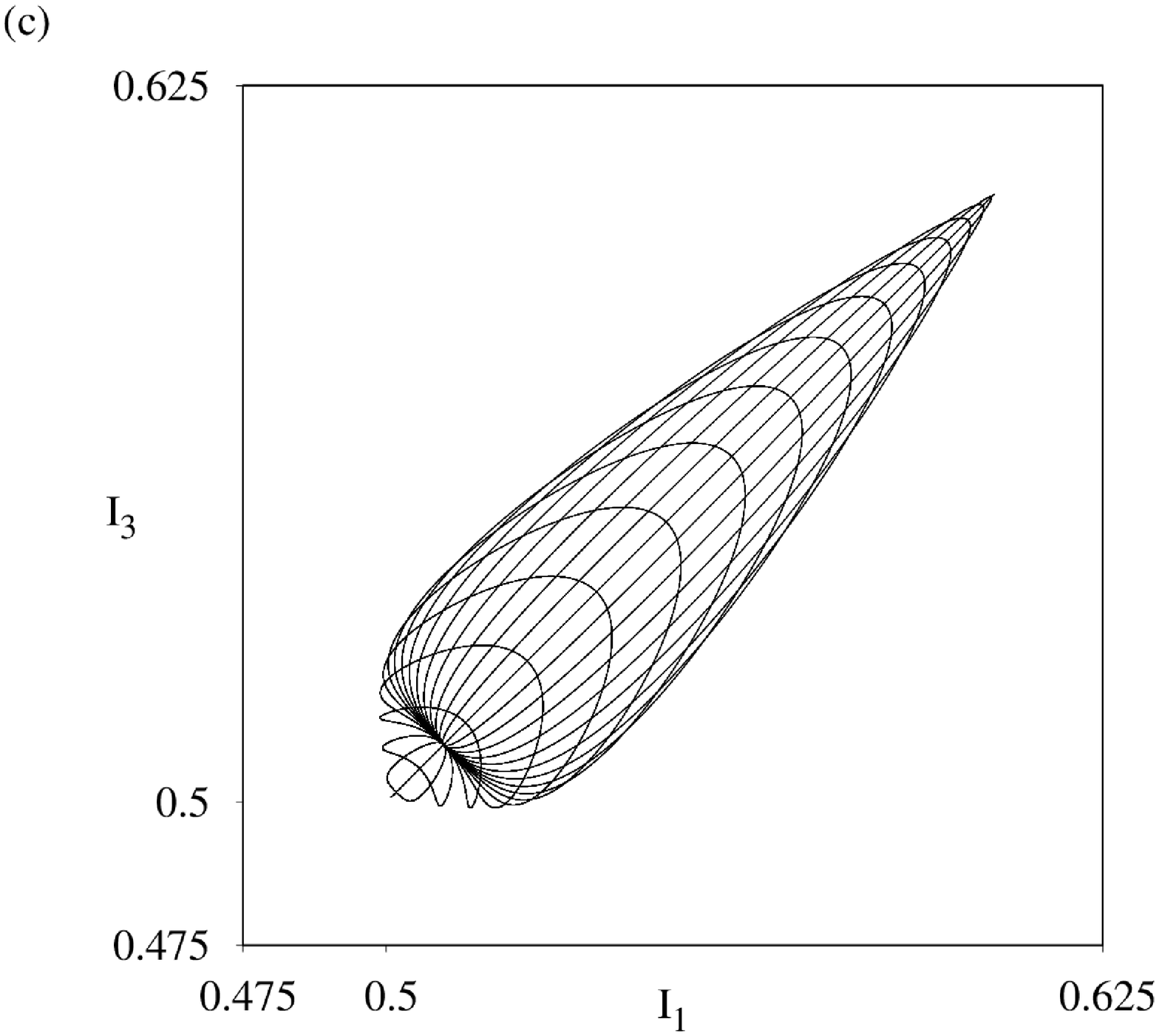}
\includegraphics[width=.5\textwidth,keepaspectratio]{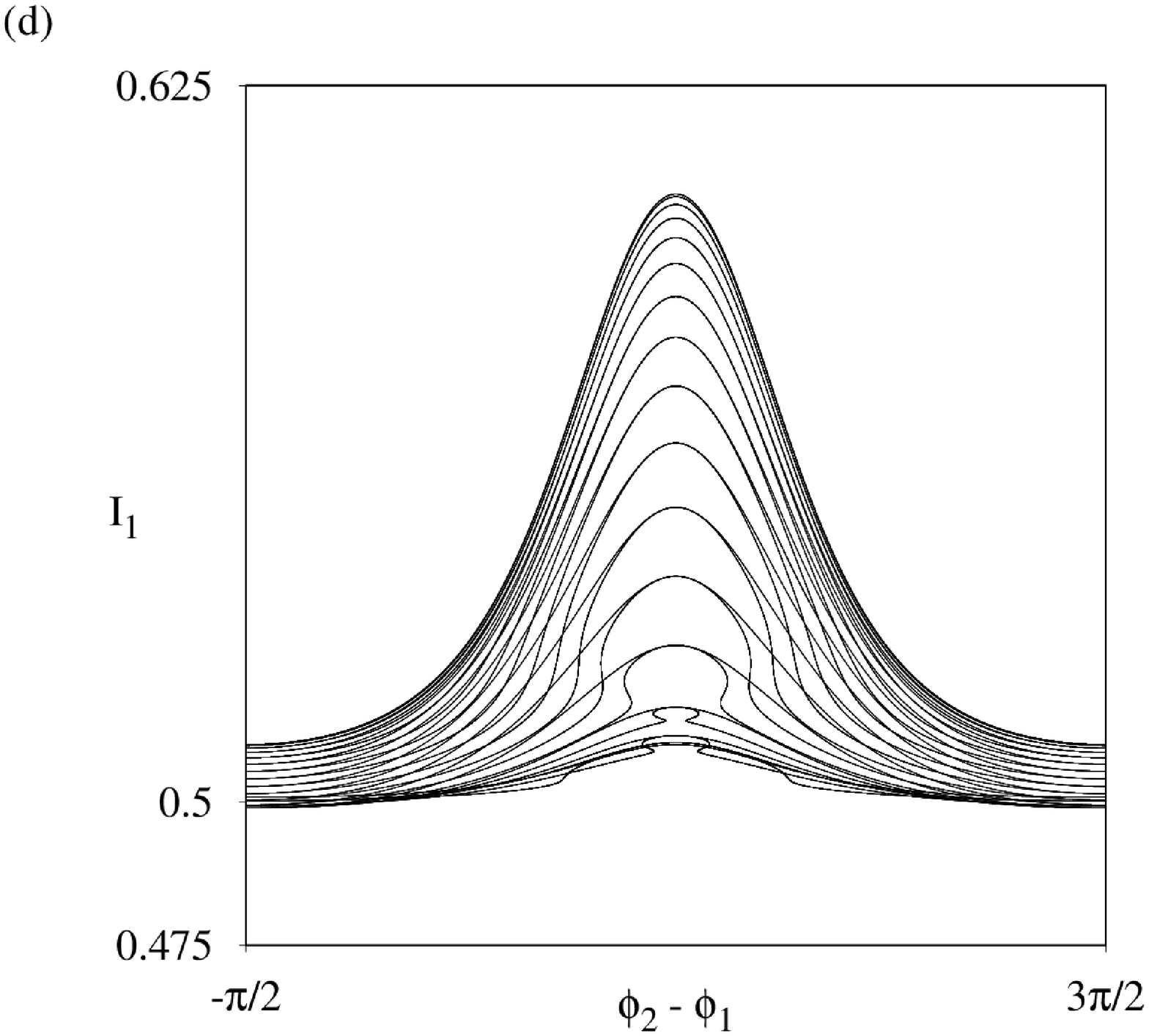}
\caption{\small
Phase portraits of system~\eqref{eq9} composed of $N=3$ oscillators at parameter values $\beta = 0$, $\varepsilon = 0.39$, $\omega_1 = - 1$, $\omega_2 = 0$, $\omega_3 = 1$. (a) Phase portrait for fixed populations $I_1=I_2=I_3=1/2$ and initial
phases for all trajectories $\phi _2 - \phi _1 = \pi - (\phi _3 - \phi _2)$. (b,c,d) Phase portraits for unfixed populations $I_1=I_3=1/2+0.01$, $I_2=1/2-0.02$, panel (b) shows dynamics of phase shifts, (c) shows evolution of populations of first and third oscillators,
(d) illustrates dynamics of population of first oscillator vs. phase shift between first and second oscillators. 
}
\label{fig01}
\end{figure}

Fig.~\ref{fig01} (a) shows the phase portrait of lattice~\eqref{eq9} composed of $N=3$ oscillators with fixed populations $I_1=I_2=I_3=1/2$ and 
initial phases for all trajectories $\phi _2 - \phi _1 = \pi - (\phi _3 - \phi _2)$. Parameters are $\beta = 0$, $\varepsilon = 0.39$, $\omega_1 = - 1$, $\omega_2 = 0$, $\omega_3 = 1$. It is a family of reversible periodic orbits on invariant torus. This result is
similar to phase model~\eqref{eq2}~\cite{1}. Fig.~\ref{fig01} (b,c,d) show different projections of phase space for lattice with unfixed populations $I_1=I_3=1/2+0.01$, $I_2=1/2-0.02$ (total population is still $C^2 = 3/2$) and 
$\phi _2 - \phi _1 = \pi - (\phi _3 - \phi _2)$. Trajectories are reversible. If phase shifts $\phi _2 - \phi _1$ and $\phi _3 - \phi _2$ are close to $\pi/2$, populations deviate greatly from uniform distribution. 

\begin{figure}[!h]
\centering
\includegraphics[width=.8\textwidth,keepaspectratio]{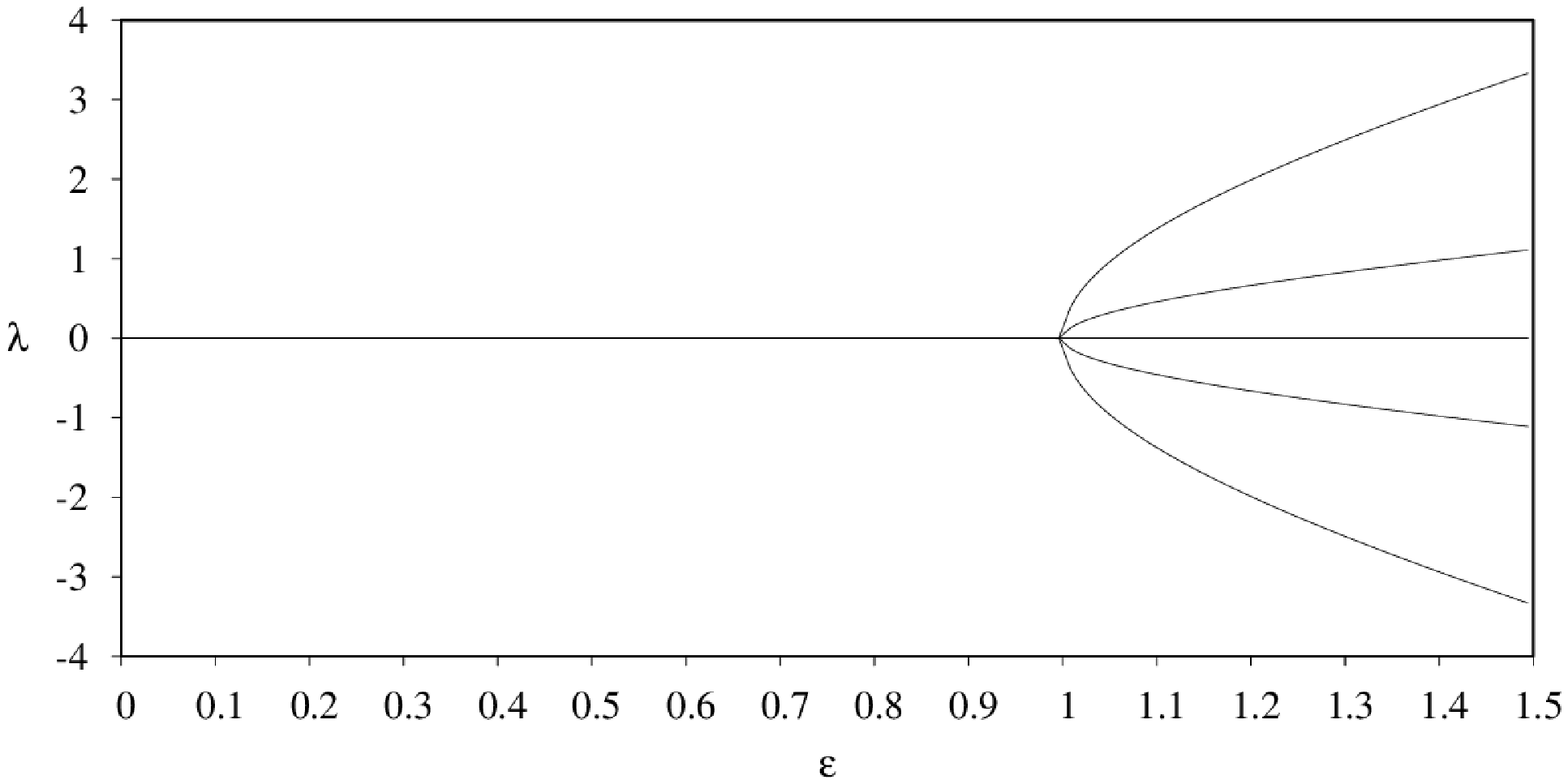}
\caption{\small
Lyapunov exponents vs. $\varepsilon$ on the invariant torus for $N=3$. $\beta = 0$, $\omega_1 = - 1$, $\omega_2 = 0$, $\omega_3 = 1$.
}
\label{fig02}
\end{figure}

Following result requires clarification. At $\varepsilon = 1$ trajectories on the two-dimensional invariant torus $I_1=I_2=I_3=1/2$ condense at $\phi _2 - \phi _1 = \phi _3 - \phi _2 = \pi/2$. At $\varepsilon > 1$ points $\mathcal{O}_1$ and $\mathcal{O}_2$ appear which
are effectively an attractor and a repeller of phase model~\eqref{eq2} acting on the invariant torus $I_1=I_2=I_3=1/2$ of model~\eqref{eq9} with fixed $\mathcal{H} = 0$ and $C^2=3/2$. It is important to note that invariant torus $I_1=I_2=I_3=1/2$ at
$\varepsilon > 1$ is actually a stable manifold of saddle equilibrium $\mathcal{O}_1$ and unstable manifold of saddle equilibrium $\mathcal{O}_2$ in four-dimensional phase space. These two points are in involution with each other (and connected by heteroclinic
trajectories) so we can discuss only one of them. Saddle equilibrium $\mathcal{O}_1$ has two-dimensional stable manifold (the invariant torus with fixed populations $I_1=I_2=I_3=1/2$) and two-dimensional unstable manifold. We
illustrate this conclusion with calculation of Lyapunov exponents of Hamiltonian model~\eqref{eq9} for trajectories with $I_1=I_2=I_3=1/2$. Fig.~\ref{fig02} demonstrates dependence of
Lyapunov exponents from $\varepsilon$. At $\varepsilon < 1$ all Lyapunov exponents are zero. This corresponds to periodic orbits on invariant torus. At $\varepsilon > 1$ only two Lyapunov exponents are zero (since we have two constants of motion), and other four are
two pairs of positive and negative exponents. This corresponds to two saddle equilibriums in involution. Stability of equilibrium $\mathcal{O}_1$ on invariant torus is compensated by instability in its neigborhood resulting in fast deviation of populations. In fact we 
can call states at $\varepsilon > 1$ phase-locked and states at $\varepsilon < 1$ amplitude-locked. We are confident that at $\varepsilon > 1$ chaotic trajectories exist outside the invariant torus $I_1=I_2=I_3=1/2$ but we have troubles with their numerical 
demonstration. Observation of attractors and repellers in conservative system~\eqref{eq9} with condition $I_1=I_2=I_3=1/2$ gives us reason to compare phase model~\eqref{eq2} with nonholonomic mechanical system. 

To visualize dynamics of system~\eqref{eq9} with more oscillators we need to construct suitable Poincar\'{e} cross-section. For the case of $N=4$ oscillators we follow~\cite{1} and choose cross-section by surface $\phi _3 - \phi _2 = \pi/2$
so that the invariant set of the involution is $\phi _2 - \phi _1 = \pi - (\phi _4 - \phi _3)$ on this surface. 

\begin{figure}[!h]
\includegraphics[width=.5\textwidth,keepaspectratio]{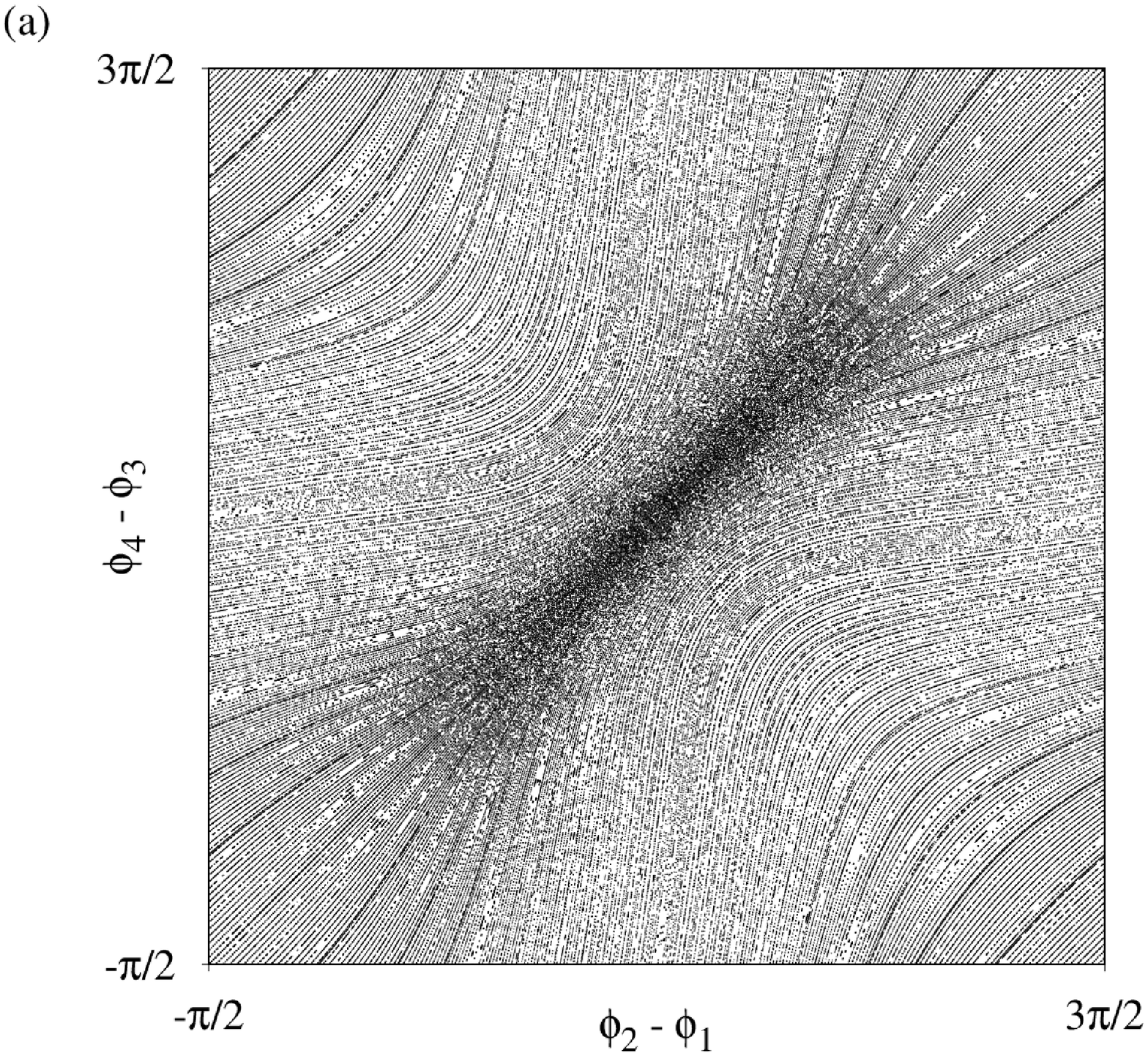}
\includegraphics[width=.5\textwidth,keepaspectratio]{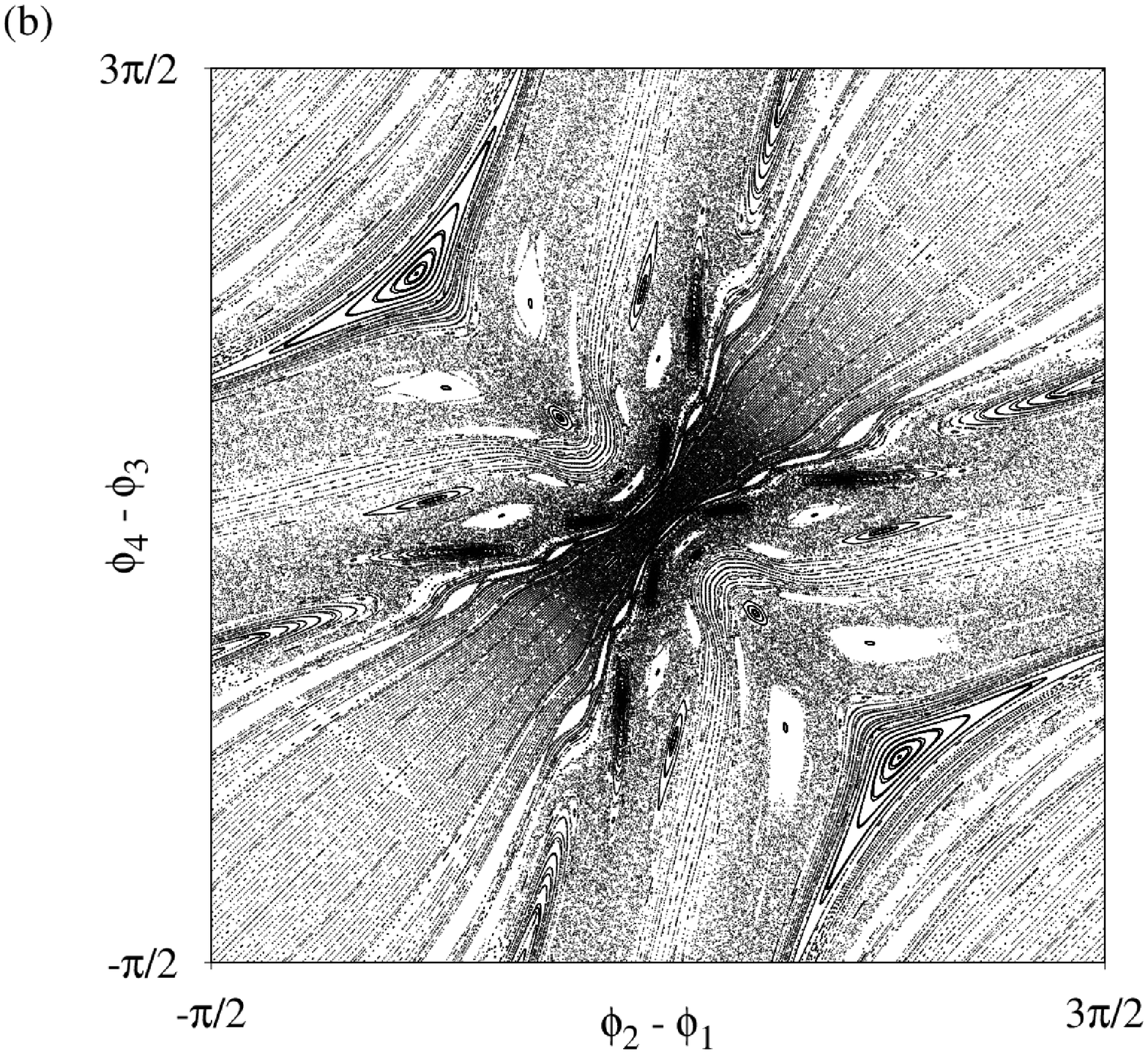}
\includegraphics[width=.5\textwidth,keepaspectratio]{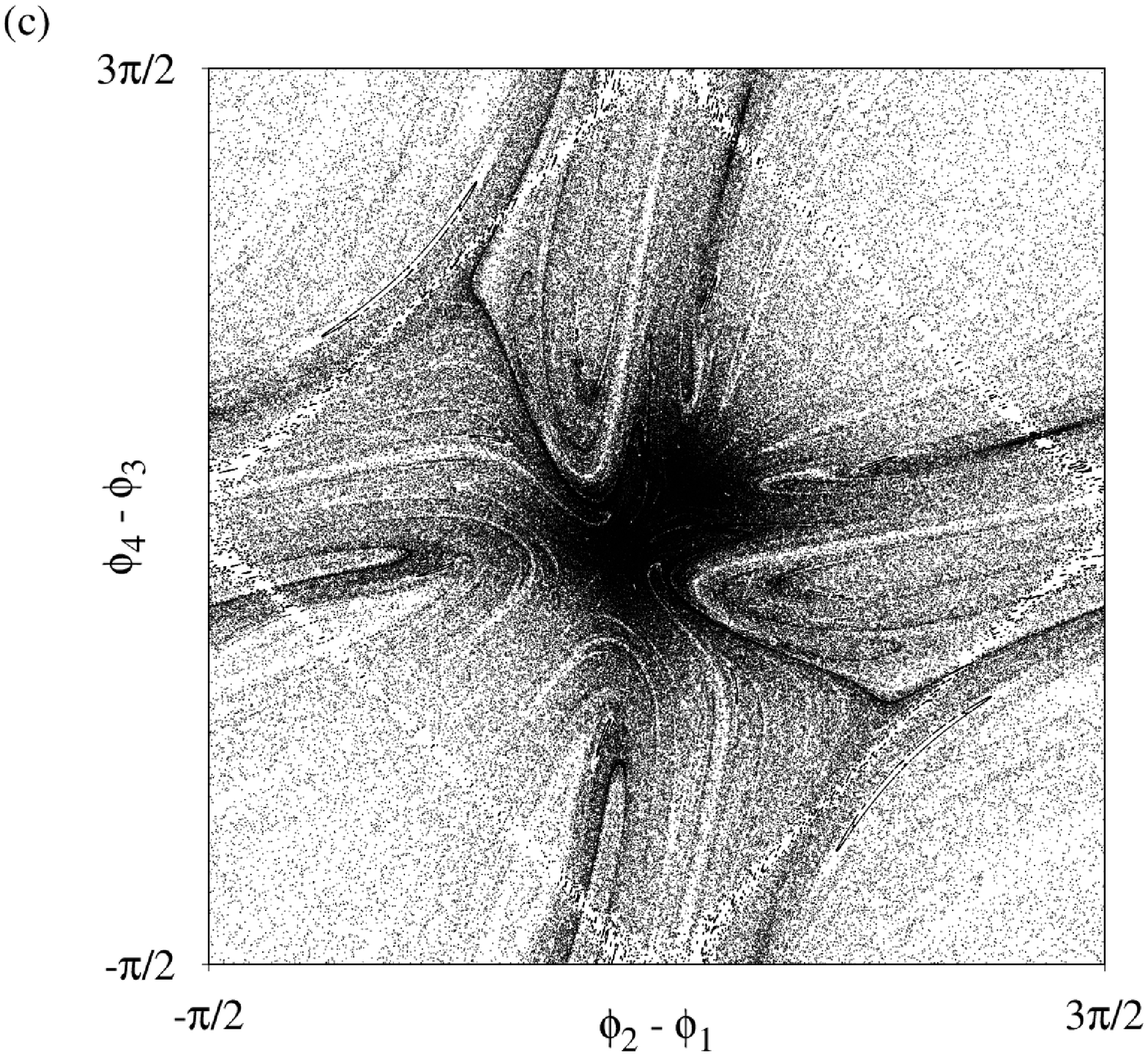}
\includegraphics[width=.5\textwidth,keepaspectratio]{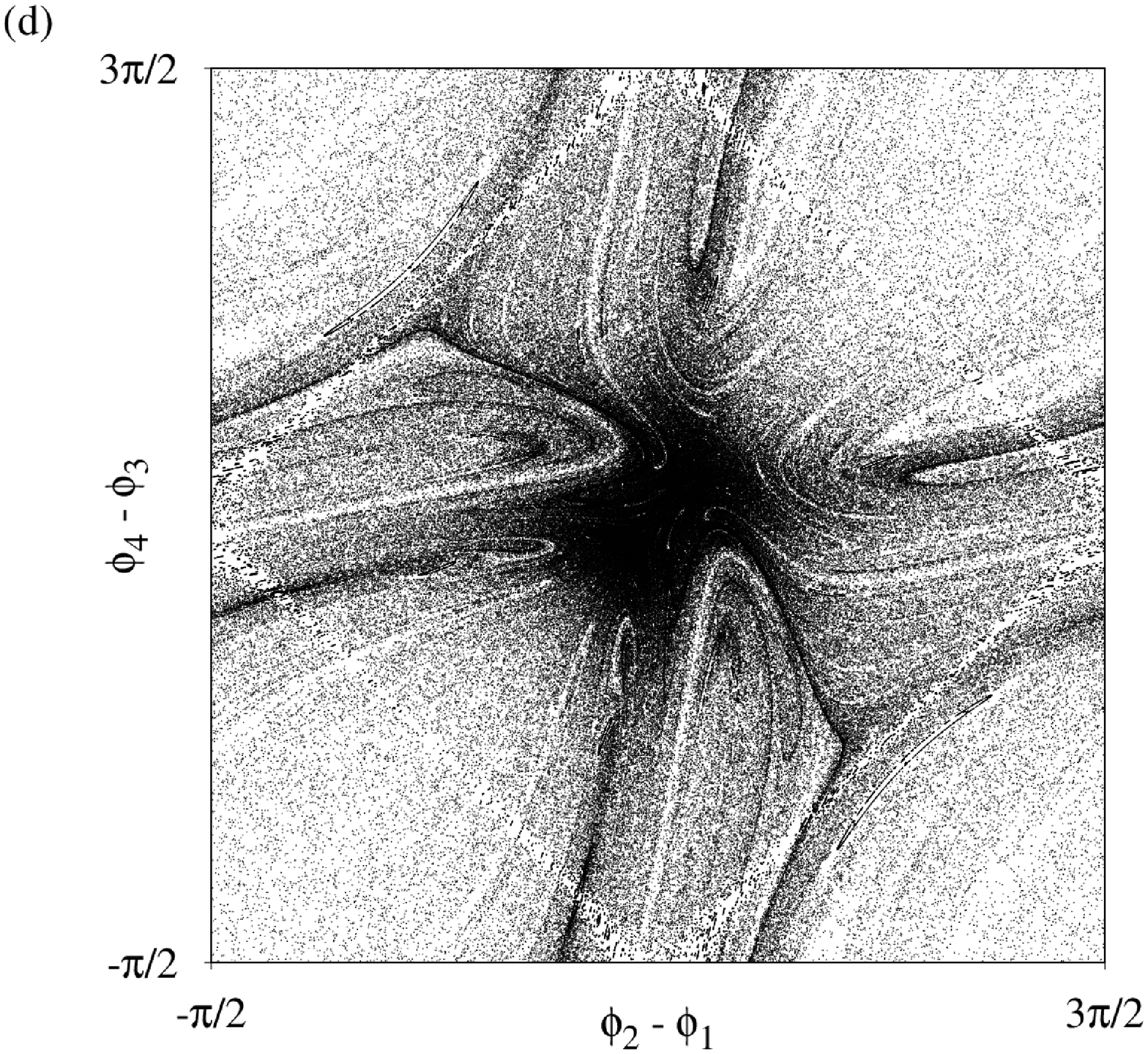}
\caption{\small
Phase portraits in Poincar\'{e} cross-section by surface $\phi _3 - \phi _2 = \pi/2$ of system~\eqref{eq9} composed of $N=4$ oscillators. Populations $I_1=I_2=I_3=I_4=1/2$ are fixed and 
initial phases for all trajectories satisfy $\phi _2 - \phi _1 = \pi - (\phi _4 - \phi _3)$, $\phi _3 - \phi _2 = \pi/2$. Parameter values $\beta = 0$, $\omega_1 = - 1.5$, $\omega_2 = -0.5$, $\omega_3 = 0.5$, $\omega_4 = 1.5$. At 
$\varepsilon = 0.19$ most of trajectories are regular (a). At $\varepsilon = 0.39$ some of trajectories are chaotic, but attractors and repellers are absent (b). At $\varepsilon = 0.49$ chaotic attractor and repeller coexist, panel (c) demonstrates evolution forward
in time, panel (d) demonstrates evolution backward in time. 
}
\label{fig03}
\end{figure}

First we discuss dynamics on the invariant torus $I_j=1/2$. It is three-dimensional in flow system~\eqref{eq9} and two-dimensional in Poincar\'{e} cross-section. 
Fig.~\ref{fig03} demonstrates phase portraits in Poincar\'{e} cross-section of lattice~\eqref{eq9} composed of $N=4$ oscillators. Populations $I_1=I_2=I_3=I_4=1/2$ are fixed and 
initial phases for all trajectories belong to the invariant set of involution. Parameters are $\beta = 0$, $\omega_1 = - 1.5$, $\omega_2 = -0.5$, $\omega_3 = 0.5$, $\omega_4 = 1.5$. 
For small coupling $\varepsilon = 0.19$ dynamics on the invariant torus $I_j=1/2$ is quasiperiodic and measure-preserving. For larger coupling $\varepsilon = 0.39$ invariant torus
$I_j=1/2$ contains quasi-periodic and chaotic trajectories, but dynamics is still measure-preserving. At coupling $\varepsilon = 0.49$ invariant torus $I_j=1/2$ has coexisting attractor $\mathcal{A}$ with its reversal repeller $\mathcal{R}$ and dynamics is no more
measure-preserving. In Poincar\'{e} cross-section attractor $\mathcal{A}$ has one-dimensional unstable manifold and one-dimensional stable manifold and phase volume on the invariant torus is overall contracting. Nevertheless in the neigborhood of the invariant torus 
$I_j=1/2$ there are two directions of expansion and contraction that compensate contraction of phase volume on invariant torus. At $\varepsilon > 0.6$ invariant torus $I_j=1/2$ has two stable and unstable periodic orbits $\mathcal{C}_1$ and $\mathcal{C}_2$ in 
involution. The stable orbit $\mathcal{C}_1$ has two-dimensional stable manifold in Poincar\'{e} cross-section. 

\begin{figure}[!h]
\centering
\includegraphics[width=.8\textwidth,keepaspectratio]{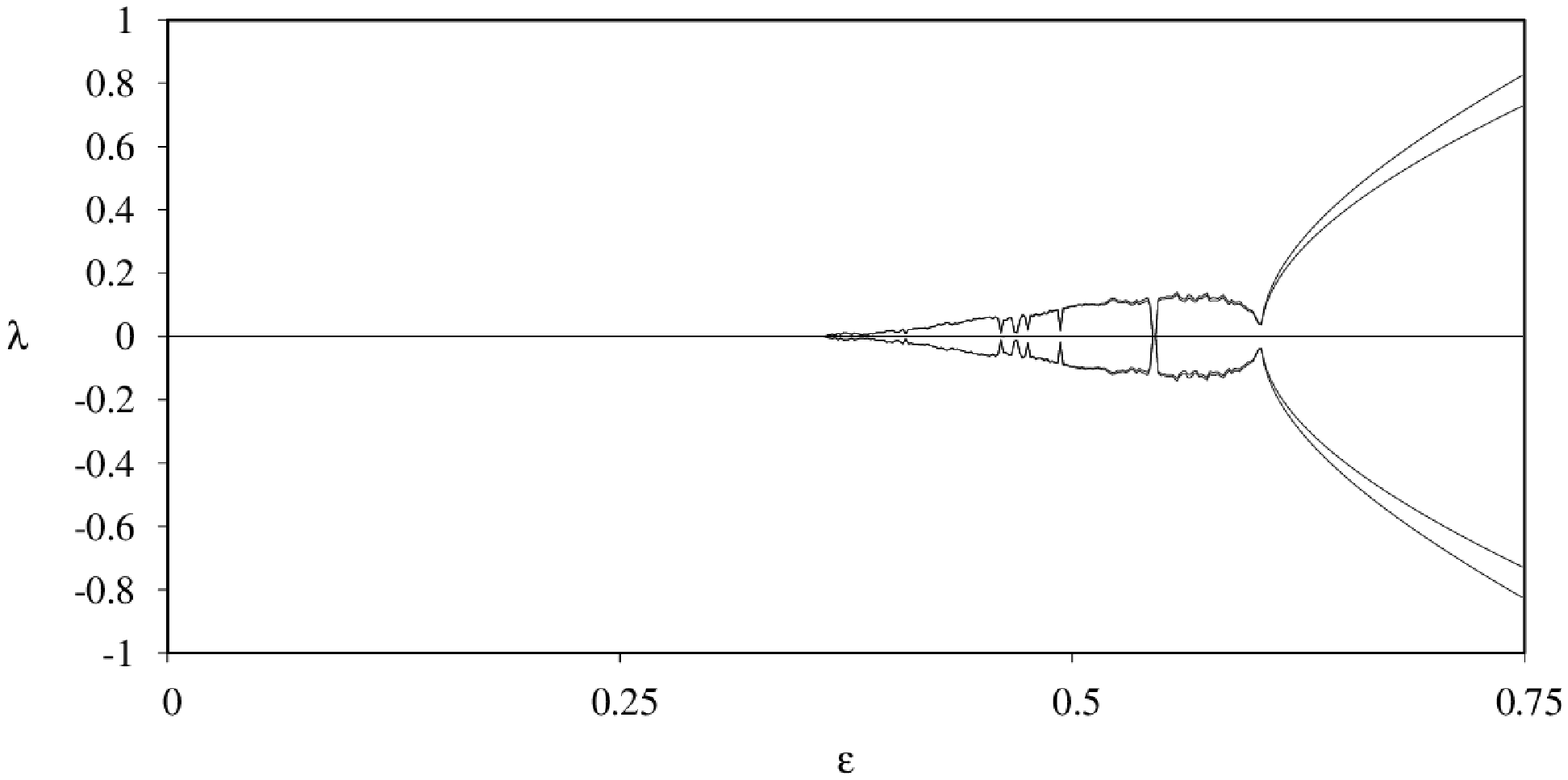}
\caption{\small
Lyapunov exponents vs. $\varepsilon$ on the invariant torus for $N=4$. $\beta = 0$, $\omega_1 = - 1.5$, $\omega_2 = -0.5$, $\omega_3 = 0.5$, $\omega_4 = 1.5$. Initial condition for all values of $\varepsilon$ was $I_1=I_2=I_3=I_4=1/2$, 
$ \phi_1 = -\pi/2$, $ \phi_2 = -\pi/3$, $ \phi_3 = \pi/6$, $ \phi_4 = \pi$. This initial condition belongs to the invariant set of involution. 
}
\label{fig04}
\end{figure}

Fig.~\ref{fig04} demonstrates dependence on $\varepsilon$ of Lyapunov exponents of Hamiltonian model~\eqref{eq9} for trajectories with $I_j=1/2$. There are eight Lyapunov exponents, but four of them are always zero (if trajectory is not an equilibrium), 
two due to constants of motion and two due to invariance to arbitrary time shift along the trajectory (one perturbation vector is tangent to the reference phase trajectory) and to arbitrary phase shift. We distinguish four regions of $\varepsilon$. 
The first corresponds to quasiperiodic motions with all exponents equal to zero ($\varepsilon < 0.36$ approximately). The second corresponds to coexistence of quasiperiodic and chaotic motions with one positive and one negative exponent for perturbations on the invariant
torus and one positive and one negative exponent for perturbations outside the invariant torus ($\varepsilon < 0.43$ according to~\cite{1}). All nonzero Lyapunov exponents are equal in magnitude. The third is the interval with chaotic attractors $\mathcal{A}$ and repellers
$\mathcal{R}$ ($\varepsilon < 0.6$) with one positive and one negative exponent on the invariant torus, nonequal in magnitude. For $\varepsilon > 0.6$ there are periodic orbits $\mathcal{C}_1$ and $\mathcal{C}_2$. Periodic orbit $\mathcal{C}_1$ has two nonequal negative 
Lyapunov exponents on the invariant torus, and two nonequal positive Lyapunov exponents outside the invariant torus. Periodic orbit $\mathcal{C}_2$ is in involution with $\mathcal{C}_1$. One can compare dependence of Lyapunov exponents for the Hamiltonian 
system~\eqref{eq9} with the dependence of Lyapunov exponents for the system~\eqref{eq2} in~\cite{1}, where exponents for perturbations outside the invariant torus are absent. 

\begin{figure}[!h]
\includegraphics[width=.5\textwidth,keepaspectratio]{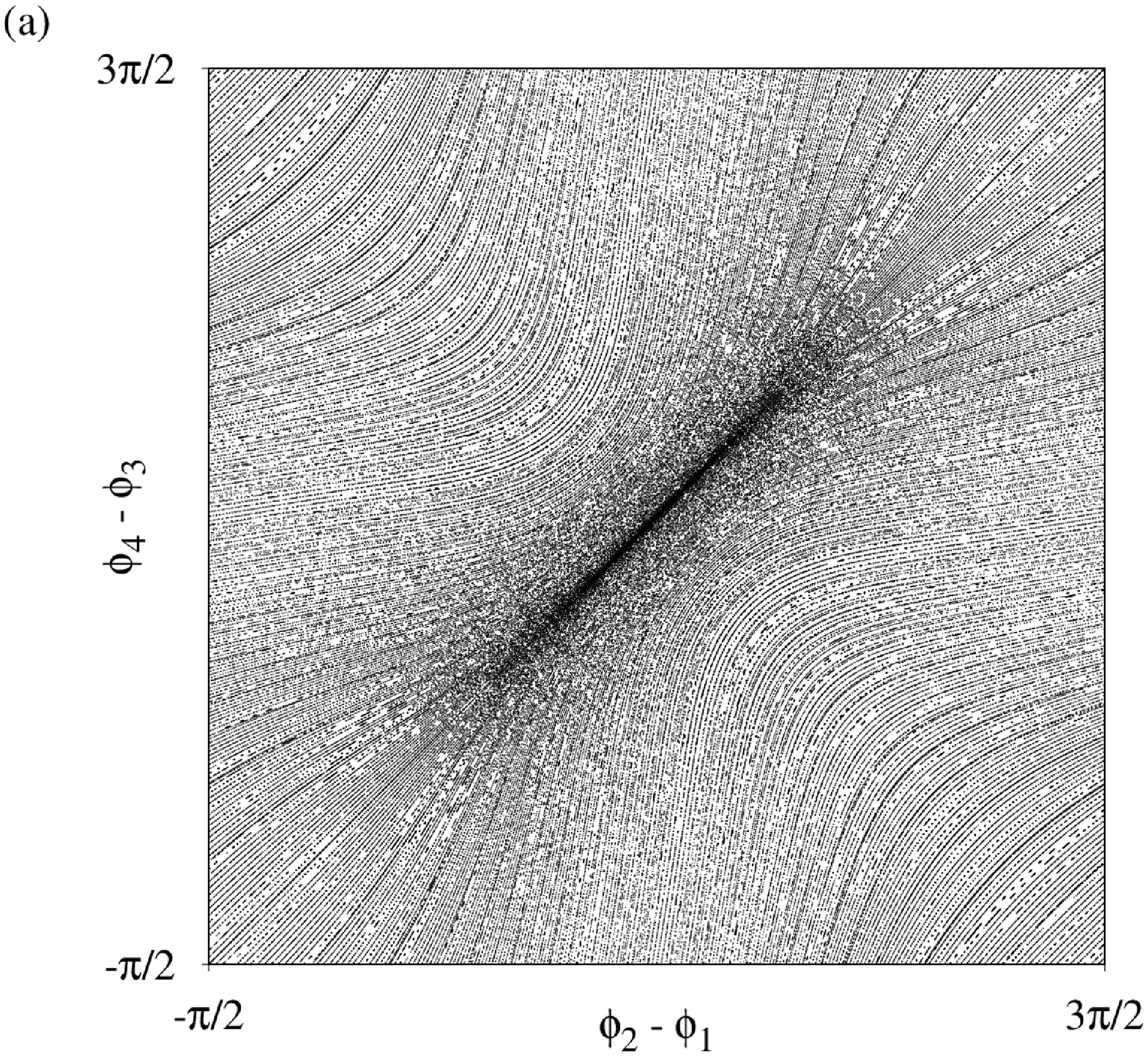}
\includegraphics[width=.5\textwidth,keepaspectratio]{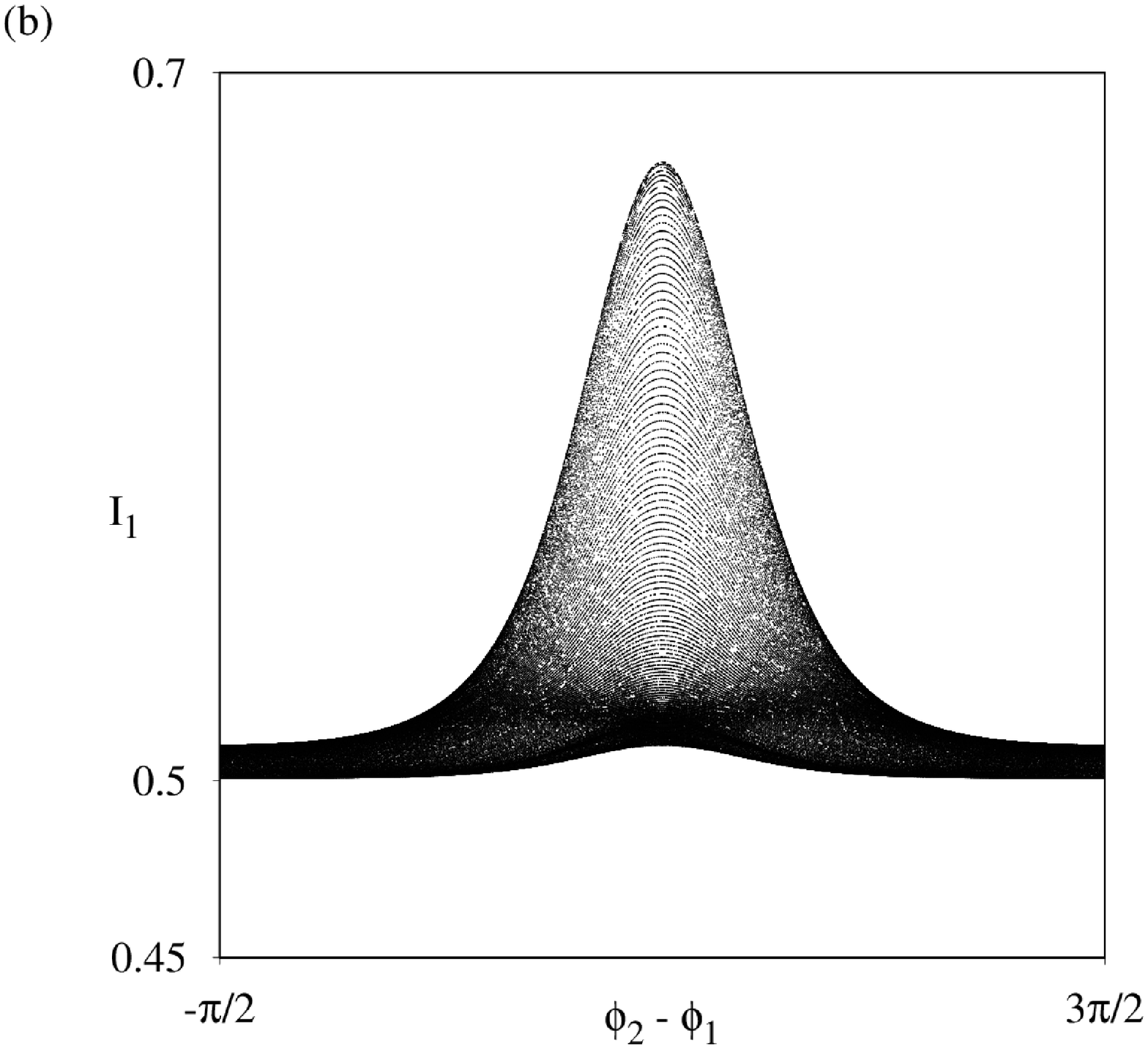}
\caption{\small
Phase portraits in Poincar\'{e} cross-section by surface $\phi _3 - \phi _2 = \pi/2$ of system~\eqref{eq9} composed of $N=4$ oscillators. Populations are not fixed. 
Initial conditions for all trajectories satisfy $I_1 = 1/2+0.01$, $I_2 = 1/2-0.01$, $I_3 = 1/2-0.01$, $I_4 = 1/2+0.01$, $\phi _2 - \phi _1 = \pi - (\phi _4 - \phi _3)$, $\phi _3 - \phi _2 = \pi/2$. Parameter values 
$\varepsilon = 0.19$, $\beta = 0$, $\omega_1 = - 1.5$, $\omega_2 = -0.5$, $\omega_3 = 0.5$, $\omega_4 = 1.5$. Panel (a) shows dynamics of phases, panel (b) shows population vs. phase shift. 
}
\label{fig05}
\end{figure}

\begin{figure}[!h]
\includegraphics[width=.5\textwidth,keepaspectratio]{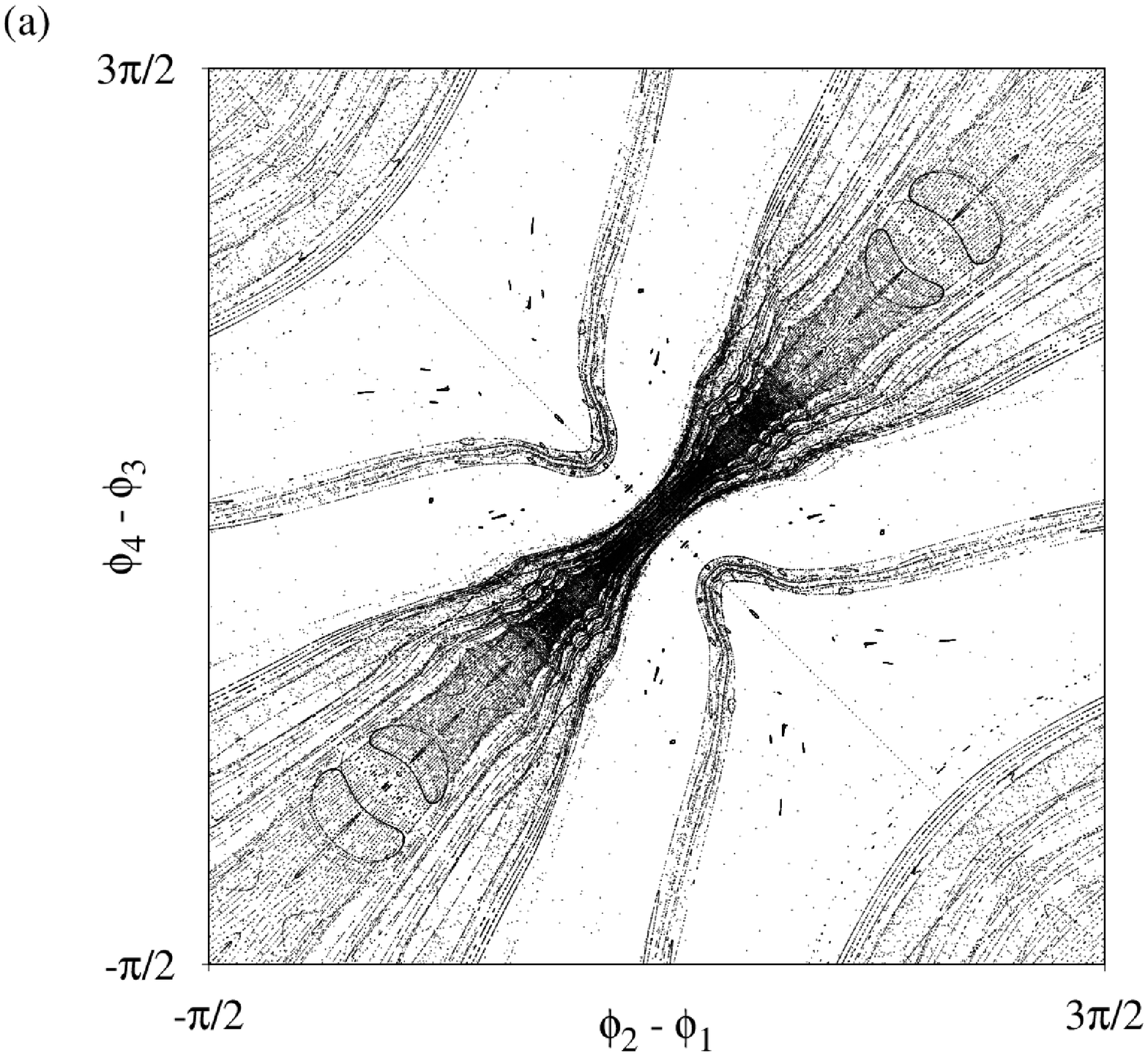}
\includegraphics[width=.5\textwidth,keepaspectratio]{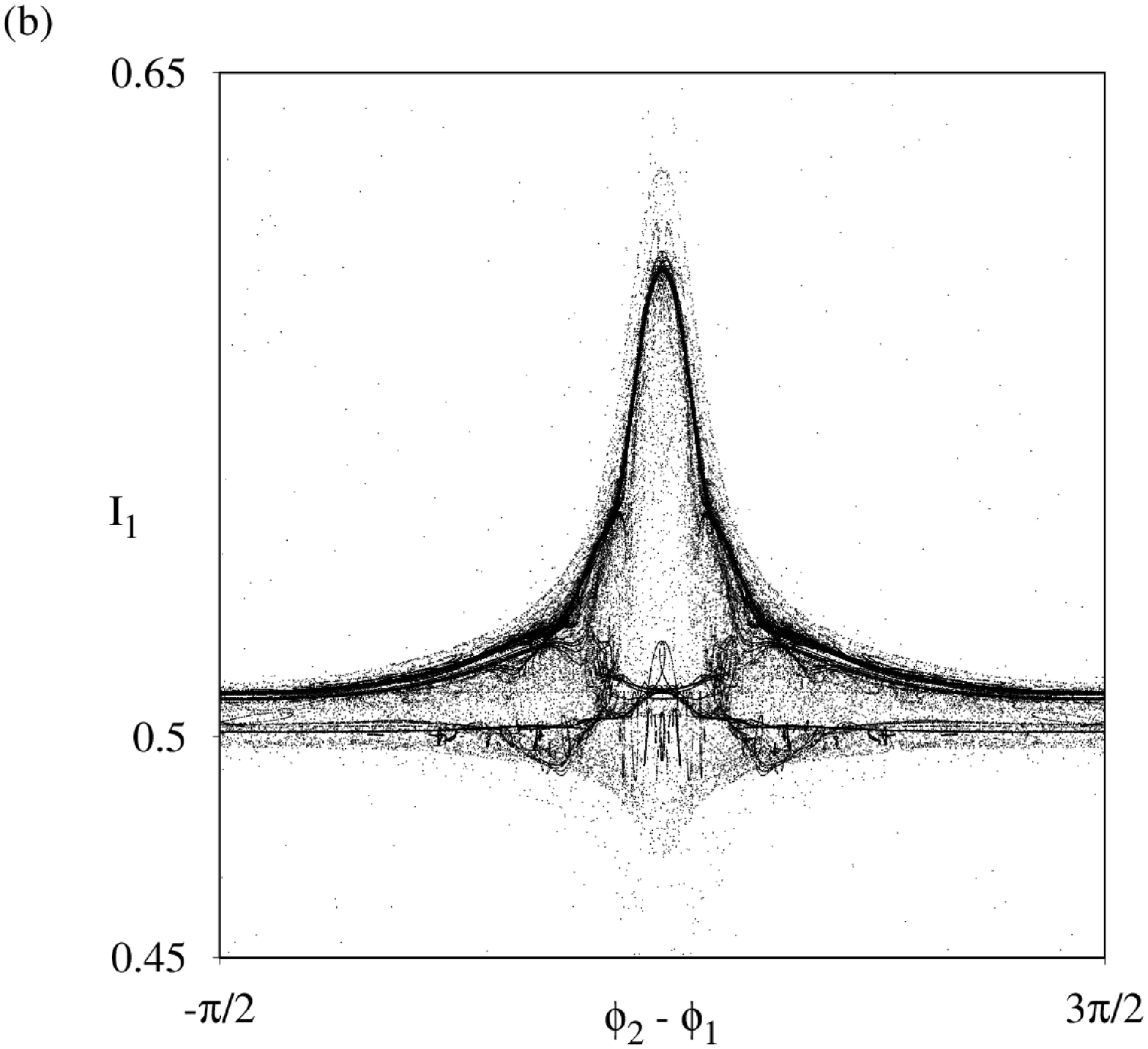}
\caption{\small
Phase portraits in Poincar\'{e} cross-section by surface $\phi _3 - \phi _2 = \pi/2$ of system~\eqref{eq9} composed of $N=4$ oscillators. Populations are not fixed. 
Initial conditions for all trajectories satisfy $I_1 = 1/2+0.01$, $I_2 = 1/2-0.01$, $I_3 = 1/2-0.01$, $I_4 = 1/2+0.01$, $\phi _2 - \phi _1 = \pi - (\phi _4 - \phi _3)$, $\phi _3 - \phi _2 = \pi/2$. Parameter values 
$\varepsilon = 0.39$, $\beta = 0$, $\omega_1 = - 1.5$, $\omega_2 = -0.5$, $\omega_3 = 0.5$, $\omega_4 = 1.5$. Panel (a) shows dynamics of phases, panel (b) shows population vs. phase shift. 
}
\label{fig06}
\end{figure}

Now we discuss motions outside the invariant torus $I_j=1/2$. Fig.~\ref{fig05} shows phase portrait in Poincar\'{e} cross-section of the lattice~\eqref{eq9} composed of $N=4$ oscillators at coupling $\varepsilon = 0.19$ with non-constant populations.  
Initial phases for all trajectories belong to the invariant set of involution. Dynamics is mostly quasiperiodic. Fig.~\ref{fig06} demonstrates phase portrait in Poincar\'{e} cross-section at coupling 
$\varepsilon = 0.39$ with non-constant populations. We describe motions in neigborhood of the invariant torus following paper~\cite{6}. If phase shift between two neiboring oscillators becomes close to $\pi/2$, their populations unlock from vicinity of invariant torus 
$I_j \approx 1/2$ and grow fast and then return. These bursts occur along the unstable manifolds of saddle sets on the invariant torus, then trajectories return to vicinity of $I_j \approx 1/2$ along the stable manifolds of saddle sets on the invariant torus. 

Fig.~\ref{fig04} shows dependence on $\varepsilon$ of Lyapunov exponents of Hamiltonian model~\eqref{eq9} for trajectories with non-constant populations. If $\varepsilon > 0.34$ four exponents become non-zero. Since dynamics is Hamiltonian, total sum 
of all Lyapunov exponents is zero. 

\begin{figure}[!h]
\centering
\includegraphics[width=.8\textwidth,keepaspectratio]{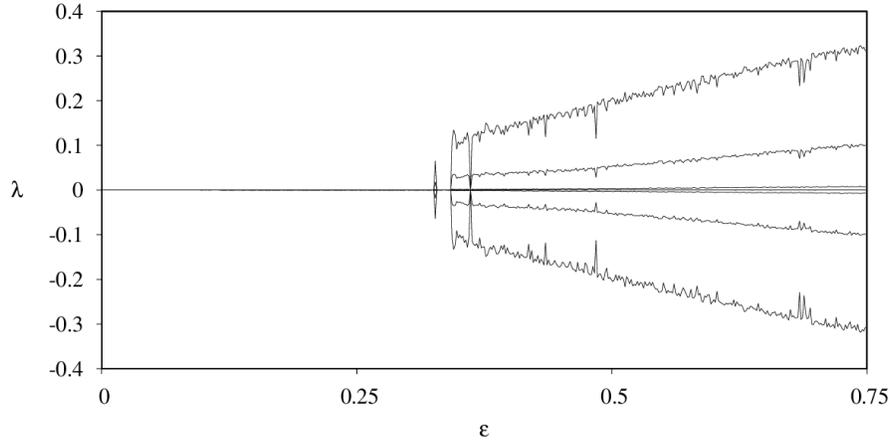}
\caption{\small
Lyapunov exponents vs. $\varepsilon$ for $N=4$ with non-constant populations. $\beta = 0$, $\omega_1 = - 1.5$, $\omega_2 = -0.5$, $\omega_3 = 0.5$, $\omega_4 = 1.5$. Initial condition for all values of $\varepsilon$ was 
$I_1 = 1/2+0.01$, $I_2 = 1/2-0.01$, $I_3 = 1/2-0.01$, $I_4 = 1/2+0.01$, 
$ \phi_1 = -\pi/2$, $ \phi_2 = -\pi/3$, $ \phi_3 = \pi/6$, $ \phi_4 = \pi$. 
}
\label{fig07}
\end{figure}

\section{Dissipative models close to Topaj -- Pikovsky system}

We propose two dissipative models close to Topaj -- Pikovsky system. The first is a lattice of dissipative pendulums with local coupling:
\begin{equation}
m \ddot{\psi} _j + \dot{\psi} _j = 1 + \varepsilon \sin \psi _{j+1}  + \varepsilon \sin \psi _{j-1}
- 2 \varepsilon \sin \psi _{j},
\label{eq13}
\end{equation}
If we set masses $m=0$, then the system~\eqref{eq13} becomes the Topaj -- Pikovsky model. Model~\eqref{eq13} with non-zero masses is not reversible. Terms with second derivatives destroy ``fat'' attractor~\cite{7} of Topaj -- Pikovsky system. If masses are small, then
 transient trajectories appear similar to Topaj -- Pikovsky phase portraits at short times of evolution. 

The second model is a lattice of locally coupled amplitude-phase equations derived from van der Pol equations:
\begin{equation}
\begin{aligned}
\dot{\alpha} _j &= \frac{1}{2} \left( \lambda - \alpha _j^2 \right) \alpha _j - \\
& - \varepsilon \left( \alpha _j - \alpha _{j-1} \cos{\left(\phi _{j-1} - \phi _j\right)} \right) 
- \varepsilon \left( \alpha _j - \alpha _{j+1} \cos{\left(\phi _{j+1} - \phi _j\right)} \right), \\ 
\dot{\phi} _j &= \omega _j + \varepsilon \frac{\alpha _{j-1}}{\alpha _j} \sin{\left(\phi _{j-1} - \phi _j\right)} 
+ \varepsilon \frac{\alpha _{j+1}}{\alpha _j} \sin{\left(\phi _{j+1} - \phi _j\right)},
\label{eq14}
\end{aligned}
\end{equation}
with $\omega _{j+1} - \omega _j = 1$.

If amplitudes $\alpha _j$ are close to constant: $\alpha _j = \sqrt{\lambda} \left( 1 + r_j \right)$, $r_j \ll 1$ then we derive
\begin{equation}
\begin{aligned}
\dot{r} _j &= - \lambda r_j
+ \varepsilon \left( \cos{\left(\phi _{j-1} - \phi _j\right)} - 1 \right) 
+ \varepsilon \left( \cos{\left(\phi _{j+1} - \phi _j\right)} - 1 \right) , \\ 
\dot{\phi} _j &= \omega _j + \varepsilon \left( 1+r_{j-1} - r_{j} \right) \sin{\left(\phi _{j-1} - \phi _j\right)} 
+ \varepsilon \left( 1+r_{j+1} - r_{j} \right) \sin{\left(\phi _{j+1} - \phi _j\right)}.
\label{eq15}
\end{aligned}
\end{equation}
Now we can reduce dimension of system~\eqref{eq15}  $\rho _j = r_{j+1} - r_j$, $\psi _j = \phi _{j+1} - \phi _j$:
\begin{equation}
\begin{aligned}
\dot{\rho} _j &= - \lambda \rho _j + \varepsilon \cos{\psi _{j+1}} - \varepsilon \cos{\psi _{j-1}}, \\ 
\dot{\psi} _j &= \Delta _j  
+ \varepsilon \left( 1 + \rho _{j+1} \right) \sin{\psi _{j+1}} 
+ \varepsilon \left( 1 - \rho _{j-1} \right) \sin{\psi _{j-1}} 
- 2 \varepsilon \sin {\psi _j}.
\label{eq16}
\end{aligned}
\end{equation}
System~\eqref{eq16} is close to Topaj -- Pikovsky model but not reversible. 

\section{Conclusion}

We discussed Hamiltonian model of oscillator lattice that describes spatial modes of one-dimensional Bose -- Einstein condensate in tilted optical lattice. Phase space of Hamiltonian model has invariant manifolds with dynamics governed exactly by Topaj -- Pikovsky
equations. System is reversible with involution similar to Topaj -- Pikovsky model. We suppose this model deserves further studying. There are promising connections with phenomenon of synchronization~\cite{6,8,9}, nonholonomic mechanics and integrability. 

\section*{Funding}
Russian Science Foundation, project No. 15-12-20035.

\bibliographystyle{unsrt}  


\end{document}